\documentstyle[12pt]{article}

\begin{document}

\title{{\Large\bf 
Conserved  Currents \\ in \\Supersymmetric \\ Quantum 
Cosmology?}
\thanks{Extended version of a report written for the 1996 Awards for Essays on 
Gravitation.}
}
\author{ {\large\sf P.V. Moniz}\thanks{e-mail: {\sf prlvm10@amtp.cam.ac.uk; 
paul.vargasmoniz@ukonline.co.uk}}
\thanks{URL: {\sf http://www.damtp.cam.ac.uk/user/prlvm10}}\\ 
Department of Applied Mathematics and Theoretical Physics\\ 
University of Cambridge\\ Silver Street, Cambridge, CB3 9EW \\ United Kingdom}
\date{DAMTP -- R96/14}

\maketitle

\vspace{-1cm}

\begin{abstract}
In this paper we investigate  whether 
conserved currents 
can be sensibly defined in supersymmetric minisuperspaces. 
Our analysis deals with  $k=+1$  FRW and Bianchi 
class--A models. Supermatter in the form of scalar supermultiplets 
is included in the former. Moreover, we restrict 
ourselves to the first-order differential equations derived 
 from the Lorentz and supersymmetry 
constraints. The 
``square-root''  structure of 
N=1 supergravity was our  motivation 
 to contemplate this  interesting    research. We 
 show that conserved currents {\em cannot} 
be adequately established except for some very simple scenarios.
Otherwise,  equations
of the type $\nabla_a  J^a = 0$ may only be obtained  from 
Wheeler-DeWitt--like equations, which are  
derived from the  supersymmetric algebra of constraints. 
Two appendices are  included. In appendix A we describe 
some interesting features of  
quantum FRW cosmologies with complex 
scalar fields when  supersymmetry is present. 
In particular, we explain how the Hartle-Hawking 
state can now be satisfactorily identified. 
In  appendix B we 
initiate 
a discussion about the retrieval 
of classical properties from 
supersymmetric quantum cosmologies.

\end{abstract}

\section{Introduction}

\indent 

 N=1 supergravity \cite{6}-\cite{8} constitutes a   ``square-root'' 
\cite{2}-\cite{4}
of
 gravity:  
in finding a physical state $\Psi$,  it is 
sufficient to  solve
the Lorentz and supersymmetry constraints of the theory. The 
algebra of constraints  then 
implies    that    $ \Psi $ will consequently  obey 
 the Hamiltonian constraints\footnote{For a review on the 
canonical quantization of supersymmetric 
minisuperspaces  see, e.g.,   
ref. \cite{rev, FPDD}.}. 
This property suggests that supersymmetry 
may induce interesting and advantageous features 
within a quantum cosmological scenario. 
In fact, 
the supersymmetry and Lorentz constraints  lead in 
many cases 
to   simple {\em first-order}  
differential equations 
in the bosonic variables (cf. ref.  \cite{A5}-\cite{A18a}).
This   contrasts with the situation in  non-supersymmetric 
 quantum cosmology:    a 
 {\it second-order}  Wheeler-DeWitt  equation
 has to be solved,  employing    specific boundary conditions
\cite{5}-\cite{13}. 
Therefore, it is quite tempting to address 
some problems of usual quantum cosmology 
from a 
supersymmetric point of view. 
In particular, the issue  of 
probability densities for  a 
quantum state $\Psi$ and conservation equations of the type 
 $\nabla_a  J^a = 0$.

As 
established 
   by C. Misner \cite{G26a}, we can derive the 
 conserved current 
$
J^a \sim \Psi^* \nabla^a \Psi - \Psi \nabla^a \Psi^*$ 
from the Wheeler-DeWitt equation 
of  superspace. 
It satisfies 
$
\nabla_a J^a = 0,
$
where $\nabla_a $ constitutes the corresponding 
covariant derivative  \cite{G26a}.
We may  associate with the current 
$J^a$ a flux across a surface $\Sigma$. In particular, 
 $\Sigma$ may be defined as the hypersurface 
of constant value of the corresponding timelike coordinate 
in a minisuperspace. Moreover, 
a conserved probability 
can then be defined from $J$ 
on the set of classical trajectories. 
However, 
this  conserved current  can  be afflicted from difficulties with  
negative probabilities  \cite{5,G26a,wh7,G26c}. 

This situation bears obvious similarities with the case of a 
scalar field $\Phi$ satisfying a Klein-Gordon 
equation \cite{Kaku}. In this case, 
the surface $\Sigma$ is usually of 
constant physical time. But the fact that $J^0\left[\Phi\right]$ 
may   be   negative
  led 
to the discovery of the Dirac equation. From Dirac's equation  
a {\em new} conserved current was derived, with the advantage of inducing 
positive definite probabilities. Subsequently, 
 the concepts of anti-particles and second quantization
 were introduced  \cite{Kaku}.
The important point to emphasize here is that the Dirac 
equation constitutes  a ``square-root'' of the 
 Klein-Gordon equation. But how far can we stretch this 
tempting analogy between, on the 
one hand,  the Klein-Gordon and  Dirac 
equations and, on the other hand, the Wheeler-DeWitt 
and the equations obtained from the supersymmetry and Lorentz 
constraints? 

Within a  standard quantum cosmological formulation \cite{5}, 
the possibility that  
$J^0$  can be positive or 
negative   merely corresponds to having 
 both expanding and collapsing classical universes. 
The flow will intersect 
a {\em generic}   
$\Sigma$ at different times. 
In other words,  $J^0$  
being 
negative
is due to a bad choice of   $\Sigma$ and does not lead necessarily 
 to a third quantization.
However, the choice 
of $\Sigma$ as a surface of constant $S$ 
within a semiclassical minisuperspace
approximation\footnote{The function $S$ that we mention in the text 
represents  an {\it approximate} solution of the 
Lorentzian Hamilton-Jacobi equation. 
In a semiclassical case 
the wave function is of the WBK form $\Psi \sim 
C e^{-W}$, where $W$ and $C$ are both complex, 
$W = I_R - iS$ and $|\nabla S| \gg 
|\nabla I_R|$. 
Subsequently, $\nabla S$ satisfies 
$\nabla \cdot J =0$ with  ~ 
$J \sim e^{-I_{R}} |C^2| \nabla S
$
\cite{5,wh7,G26c}.}
is quite 
satisfactory: 
the flow associated with $\nabla S$   intersects them 
once and only once.

The objective of this paper (see also ref. \cite{Rerum}) is 
precisely 
to 
investigate if 
positive definite  \cite{Kaku} 
 conserved currents 
can be  defined in supersymmetric quantum cosmology. Namely, 
in the sense 
of those retrieved from the Dirac equation in 
standard quantum field theory.
If supersymmetric conserved currents 
 could be obtained, then  the $J^0$ negative values 
in non-supersymmetric quantum cosmology would be 
seen from a whole new perspective. In particular, maybe the presence of  supersymmetry could induce conserved currents which would 
correspond in standard quantum cosmology  to ``select'' 
 appropriate trajectories and hypersurfaces $\Sigma$ in minisuperspace.

The approach that we employ here is based on a differential 
operator representation for the  fermionic variables.
This  constitutes the correct  procedure.  In fact, 
it is totally 
 consistent with the existence of second--class constraints  
and subsequent  Dirac brackets in 
supergravity theories. 
These then imply that   fermionic variables and their Hermitian conjugates are
 intertwined within  a canonical coordinate--momentum relation
(see ref. \cite{4}-\cite{A18a}, \cite{A21,A20}  
 for further details). 

It should be pointed out that other 
authors have persued objectives similar to ours 
\cite{A19,A17a,OOcc}. However, their 
approaches involved  ``square-root'' formalisms of gravity 
distinct from the one we employ here. In particular, 
{\it rigid} 
 supersymmetry 
was used in ref. \cite{A19}. The approach present in \cite{A17a} is {\it not} 
supersymmetric.  Furthermore,  a  wave function 
arranged as a vector was 
used in ref. \cite{OOcc}-\cite{A4}, 
with a bewildering interpretation   of 
universe-anti-universe  states \cite{A4}. 
In  ref. \cite{A17a}-\cite{A4} 
 the crucial role of   the 
Lorentz constraints does not seem to have been properly dealt with or 
is even absent. 

This  paper is then organized as follows. In section 2 
we analyse a $k=+1$    FRW model in the 
framework of N=1 supergravity with a scalar 
supermultiplet \cite{A9}-\cite{A11}, \cite{A16}-\cite{A18}, 
\cite{A19a}-\cite{A18a}. Some improvements concerning 
the results present in 
\cite{A16}-\cite{A18}, 
\cite{A19a,A23}
 are  included. We also   point to 
specific differences between 
our FRW model and the ones analysed in ref. \cite{A17a,Khala}. 
The analysis of these  differences subsequently 
assist us in establishing if (Dirac-like \cite{Kaku})
  conserved currents are allowed in supersymmetric quantum cosmology. 
In section 3 we consider Bianchi 
class--A models obtained  within  pure N=1 supergravity 
\cite{A5}-\cite{A19}, \cite{A21,A20}. Some results previously  presented in the 
literature are rectified. In addition, we describe 
how the presence of   anisotropy 
prevent us from establishing  generic conservation
equations. Anisotropy and matter 
lead in general to a mixing 
of  fermion (Grassman-valued) sectors in $\Psi$. 
This is also a direct consequence of  the 
Lorentz invariance of the theory, which is 
absent in ref. \cite{A1}-\cite{A4}. 
As a result, the first-order differential equations 
derived from the supersymmetry constraints 
 become coupled.
Only the use of subsequent Wheeler-DeWitt--type equations 
can provide  consistent physical solutions. 
But in doing so we are effectively  placing  
  any discussion of conserved currents 
and probability densities {\em back} in the context of usual 
quantum cosmology \cite{A20a}.
  Our conclusions and discussions are presented in section 4.
In appendix A we present  some 
 features of $k+1$ FRW cosmologies with complex 
scalar fields that  may be  relevant in  
supersymmetric quantum cosmology. 
Finally,  we 
initiate in appendix B 
a discussion on the   topic of   retrieving 
classical properties from 
supersymmetric quantum cosmological models.

\section{Supersymmetric  FRW $k=+1$ models}

\indent

Let us consider  
 the action of the more general theory of N=1
 supergravity in the 
presence of gauged supermatter 
(see eq. (25.12) in ref. \cite{8}). 
Our physical  variables   include the tetrad 
 $ e^{A A'}_{~~~~\mu} $ (in  2-component  
spinorial form)
  and  the   gravitinos which are represented by 
$ 
 \psi^A_{~~\mu}, \bar\psi^{A'}_{~~\mu}$. 
The ``overline'' denotes Hermitian 
conjugation. 
The tetrad for  a $k=+1$ FRW 
model   can be   be written as 
 \begin{equation}
 e_{a\mu} = \left(\begin{array}{cc} 
N (\tau) &0 \\
0  &a E_{\hat a i} \end{array}\right)~,
 e^{a \mu} = \left( \begin{array}{cc}
N (\tau)^{-1} & 0 \\
0 &a (\tau)^{-1} E^{\hat a i} \end{array} \right) ~,
\label{eq:2.2}
\end{equation}
where $ \hat a $ and $ i $ run from 1 to 3 and 
$ E_{\hat a i} $ is a basis of left-invariant 1-forms on the unit $ S^3 $
with volume $ \sigma^2 = 2 \pi^2 $.
This ansatz reduces the number of degrees of freedom provided by $ e_{AA'
\mu} $. Hence, a consistent  
ansatz for $ \psi^A_{~~\mu} $ and $ \bar\psi^{A'}_{~~\mu} $
is required. We take  
$\psi^A_{~~0} $ and $ \bar\psi^{A'}_{~~0} $ to be  
 functions of time only and 
 \begin{equation} 
\psi^A_{~~i} = e^{AA'}_{~~~~i} \bar\psi_{A'}~, ~
\bar\psi^{A'}_{~~i} = e^{AA'}_{~~~~i} \psi_A~. 
\label{eq:2.3}
\end{equation}
We have  introduced  the new spinors $ \psi_A $ and their 
Hermitian  conjugate, $\bar\psi_{A'} $, which
are also functions of time only  \cite{rev}, \cite{A8}-\cite{A11},
\cite{A18a}.  
The scalar supermultiplet present in the action 
will consist of   spatially homogeneous 
complex  scalar 
fields $ \phi, \bar\phi $ and their spin-$\frac{1}{2}$ 
partners $ \chi_A (t) , \bar\chi_{A'} (t)$.
Any vector field and 
supersymmetric partners  are taken henceforth to be zero.
Moreover, we choose 
 a two-dimensional 
spherically symmetric K\"ahler geometry\footnote{For a flat 
two-dimensional  K\"ahler geometry
   we will   basically get the same 
physical information \cite{A9}-\cite{A11}, 
\cite{A16}-\cite{A17}, \cite{A19a,A23}.}.

Using the Ans\"atze previously described, 
the action of the full theory 
is  reduced to one with a 
 finite number of degrees of freedom.
The analysis 
of the reduced theory 
becomes simpler if we   redefine the  fermionic fields, $ \chi_{A} $,  
$ \psi_{A} $ as follows. 
First, we 
take  
\begin{equation}
 \hat \chi_{A} = {\sigma a^{3 \over 2} \over 
2^{1 \over 4} (1 + \phi \bar \phi)} \chi_{A}, ~ 
\hat{ \bar \chi}_{A'} = {\sigma a^{3 \over 2} 
\over 2^{1 \over 4} 
(1 + \phi \bar \phi)} \bar \chi_{A'}
\end{equation}
and   
\begin{equation}
 \hat \psi_{A} = {\sqrt{3} \over 
2^{1 \over 4}} \sigma a^{3 \over 2} \psi_{A}, 
~\hat{\bar \psi}_{A'} = {\sqrt{3} \over 2^{1 \over 4}} 
\sigma a^{3 \over 2} \bar \psi_{A'}.
\end{equation}
In addition, we  use  
 unprimed spinors, and, to this end, we define
\begin{equation} \bar \psi_{A} = 2 n_{A}^{~B'} \bar \psi_{B'}~, ~
 \bar \chi_{A} = 2 n_{A}^{~B'} \bar \chi_{B'}.
\label{eq:2.9}
 \end{equation}
These redifinitions allow for simple Dirac brackets to be 
obtained \cite{rev,A16,A22,A17}, namely 
\begin{equation}
 [\chi_{A}, \bar \chi_{B}]_{D} = -i \epsilon_{AB}~, ~
 [\psi_{A}, \bar \psi_{B}]_{D} = i \epsilon_{AB}.  
\end{equation}
Furthermore,
\begin{equation}
 [a , \pi_{a}]_{D} = 1~, ~ [\phi, \pi_{\phi}]_{D} = 1~,
 ~[\bar \phi, \pi_{\bar \phi}]_{D} = 1,
\end{equation} 
and the rest of the brackets are zero.
 At this point
 we choose $ (\chi_{A} , \psi_{A} , a , \phi , \bar \phi) $ to be the coordinates of the 
configuration space and 
$ (\bar \chi_{A} , \bar \psi_{A}$, 
  $\pi_{a}$ , $\pi_{\phi}$ , $\pi_{\bar \phi}) $ to be the momentum operators in this 
representation.
Hence, quantum mechanically we may take (with $\hbar =1$)  
\begin{equation}  
 \bar \chi^{A}   \rightarrow 
-{  \partial   \over   \partial   \chi^{A}}, \bar \psi_{A} \rightarrow {  \partial   \over   
\partial   \psi^{A}},     \pi_{a}  \rightarrow  {  \partial   \over   \partial   a}, 
\pi_{\phi} \rightarrow -i {  \partial   \over   \partial   \phi}, 
 \pi_{\bar \phi} \rightarrow -i {  \partial   \over 
  \partial   \phi}.
\end{equation} 
  Implementing all these redefinitions, the 
 supersymmetry constraints  have the differential operator form

\begin{eqnarray} 
 S_{A} & = & 
 -{i \over \sqrt{2}} (1 + \phi \bar \phi) \chi_{A} 
{  \partial   \over    \partial   \phi}
 - {1 \over 2 \sqrt{6}} a \psi_{A} {  \partial   \over  
 \partial   a} -
  \sqrt{3 \over 2} \sigma^{2}a^{2} \psi_{A} 
\nonumber \\ &-& {5i \over 4 \sqrt{2}} \bar \phi \chi_{A} \chi^{B} { 
 \partial   \over   \partial   \chi^{B}} -
 {1 \over 8 \sqrt{6}} \psi_{B} \psi^{B} {  \partial  
 \over   \partial   \psi^{A}} 
- 
{i \over 4 \sqrt{2}} \bar \phi \chi_{A} \psi^{B} {  \partial   \over   \partial   \psi^{B}}  \nonumber \\ 
& -  & {5 \over 4 \sqrt{6}} \chi_{A} \psi^{B}{  
\partial   \over   \partial   \chi^{B}}
 + {\sqrt{3} \over 4 \sqrt{2}} \chi^{B} \psi_{B} {  
\partial   \over   \partial   \chi^{A}}  
+ {1 \over 2 \sqrt{6}} \psi_{A} \chi^{B} {  \partial  
 \over   \partial   \chi^{B}},  
\label{eq:2.13}
   \end{eqnarray} 
together with  its Hermitian conjugate,  obviously using eq. (\ref{eq:2.9}). 
The   Lorentz 
constraints take the form 
  \begin{equation} 
 J_{AB} = \psi_{(A} \bar{\psi}_{B)} - \chi_{(A} \bar{\chi}_{B)}
 =  0~,
\label{eq:2.14}
\end{equation}
which   implies that  the 
 most general form for the wave function 
of the universe is 

\begin{equation} 
  \Psi = A  +  B \psi^{C} \psi_{C} + C \psi^{C} \chi_{C} + 
D \chi^{C} \chi_{C} + E \psi^{C} \psi_{C} \chi^{D} \chi_{D}, 
\label{eq:2.15}
\end{equation}
where $A$, $B$, $C$, $D$, $E$  are functions of $a$, $\phi$ ,$\bar \phi$, 
  only.

 Let us now see which physical states can be derived 
from the Lorentz and supersymmetry constraints. 
  From the constraint 
(\ref{eq:2.13}), its Hermitian conjugate 
and eq. (\ref{eq:2.15}) 
we will get four equations from 
 $ S_{A} \Psi = 0 $ and  another four equations from $ \bar S_{A'} \Psi = 0 $:

\begin{eqnarray} 
  (1 + \phi \bar \phi) {  \partial   A \over   \partial   \phi}  =   0~, 
(1 + \phi \bar \phi) {  \partial   E \over   \partial   \bar \phi} & =&  0~,  
\label{eq:2.2.1} \\ 
 { a \over 2 \sqrt{6}} {  \partial   A \over   \partial   a} + \sqrt{3 \over 2} \sigma^{2} a^{2} A  = 0~,  
  {a \over \sqrt{6}} {   \partial   E \over   \partial   a}  - 
 \sqrt{6} \sigma^{2} a^{2} E & = &  0~, \label{eq:2.2.4}
\end{eqnarray}
\begin{eqnarray}
   (1 + \phi \bar \phi) {   \partial   B \over   \partial   \phi}  + 
  {1 \over 2} \bar \phi B 
+ { a \over 4 \sqrt{3}} {  \partial   C \over   \partial   a} - {7 \over 4 \sqrt{3}} C + 
{\sqrt{3} \over 2}  \sigma^{2} a^{2} C & = &  0~, \label{eq:2.2.5} \\
{a \over \sqrt{3}} {  \partial   B \over   \partial   a}  - 
 2 \sqrt{3} \sigma^{2} a^{2} B - \sqrt{3} B
 + (1 + \phi \bar \phi) {  \partial   C \over   \partial   \bar \phi} + {3 \over 2} \phi C 
&=& 0 
~,\label{eq:2.2.6} \\ 
{a \over \sqrt{3}} {  \partial   D \over   \partial   a} 
+   2 \sqrt{3} \sigma^{2} a^{2} D - \sqrt{3} D 
- (1 + \phi \bar \phi) { \partial C \over \partial \phi} - {3 \over 2} \bar \phi C & =  &
0~, \label{eq:2.2.7} \\ 
(1 + \phi \bar \phi) {  \partial   D \over   \partial   \bar \phi}  + 
  {1 \over 2} \phi D 
- {a \over 4 \sqrt{3}} {  \partial   C \over   \partial   a} + {7 \over 4 \sqrt{3}} C + {\sqrt{3} \over 2} \sigma^{2} a^{2} C & = &  0 ~. 
\label{eq:2.2.8}
\end{eqnarray}
We can see that 
(\ref{eq:2.2.1}), 
(\ref{eq:2.2.4})  
constitute  decoupled equations for $A$ and $E$. 
 Eq. (\ref{eq:2.2.5}) and 
(\ref{eq:2.2.6}) constitute 
  coupled equations between $B$ and $C$, while  eq. (\ref{eq:2.2.7}), 
(\ref{eq:2.2.8})  
are coupled equations between $C$ and $D$. These equations 
can be decoupled employing  
 $ B = \tilde B (1 + \phi \bar \phi)^{- {1 \over 2}}$, 
$ C = {\tilde C  \over \sqrt{3}}(1 + \phi \bar \phi)^{- {3 \over 2}}$, 
 $ D = \tilde D (1 + \phi \bar \phi)^{- {1 \over 2}}$.
We can then eliminate $\tilde B$ and $\tilde D$ to get 
two  partial differential equations which imply that  
 $C = 0$ (cf. ref. \cite{A16,A22}). 

Let us multiply the first eq. in (\ref{eq:2.2.4}) by E, then multiply the 
second  by A. Their addition results in  
\begin{equation}
\frac{\partial \left(A\cdot E\right)}{\partial \, a} = 0 ~.
\label{eq:2.2.20a}
\end{equation}
Eq. (\ref{eq:2.2.20a}) seems to suggest a relation vaguely 
similar to 
$\nabla \cdot J = 0$. For the case of pure N=1 supergravity, 
eq. (\ref{eq:2.2.1})-(\ref{eq:2.2.8}) are reduced to  
just eq.  
(\ref{eq:2.2.4}). Hence, eq. (\ref{eq:2.2.20a}) 
would constitute a (very simple) conservation--type  equation 
obtained {\it directly} from the supersymmetry and Lorentz 
constraints.

It is interesting to 
notice that  eq. (\ref{eq:2.2.1})-(\ref{eq:2.2.4}) imply
\begin{equation}
A = f(\bar\phi) e^{-3\sigma^2 a^2} ~, E =  g(\phi) e^{3\sigma^2 a^2} ~, 
\label{eq:2.2.20ba}
\end{equation}
where $f, g$ are anti-holomorphic and holomorphic 
functions of $\phi, \bar\phi$, respectively. 
It seems unsatisfactory that we cannot obtain 
from 
 the Lorentz and supersymmetry constraints 
the 
explicit dependence of $\Psi$ on $\phi, \bar\phi$
(see ref. \cite{A9}-\cite{A10}, \cite{A16}-\cite{A18}, \cite{A19a,A23}). 
However, this apparent drawback 
can be circumvented as follows. Basically, 
we  introduce the variables 
$ r^2 = \phi\bar\phi$ with  $\phi = r e^{i\theta}$.
These new variables effectively decouple 
the 
two degrees of freedom
associated with $\phi$ and $\bar \phi$. 
But more importantly, they will assist us in 
establishing {\em if} generic 
conserved currents can  be defined in a 
supersymmetric
$k=+1$ 
FRW  minisuperspace
with complex 
scalar fields.

In fact, equations (\ref{eq:2.2.1}) 
can then be written as 
\begin{eqnarray}
\frac{\partial A}{\partial r}   -  
 i\frac{1}{r}\frac{\partial A}{\partial \theta} &= &0~,~\label{eq:2.2.20d}\\
\frac{\partial E}{\partial r} +   
 i\frac{1}{r}\frac{\partial E}{\partial \theta} &=& 0~. \label{eq:2.2.20e}
\end{eqnarray}
Multiply eq. (\ref{eq:2.2.20d})   by $E$ and eq. (\ref{eq:2.2.20e})
  by $A$. 
It follows from their subtraction    that
\begin{equation}
\frac{\partial (A\cdot E)}{\partial \theta } - ir  \left(
\frac{\partial E}{\partial r} A - \frac{\partial A}{\partial r} E\right) = 0.
\label{eq:2.2.20ff}
\end{equation}

Let us now 
take  $C=0$.  
 Multiply eq. (\ref{eq:2.2.6})  by $D$ and eq. 
(\ref{eq:2.2.7}) 
 by $B$. Adding them,   we get  a relation similar to (\ref{eq:2.2.20a})
\begin{equation}
D_a ( B \cdot D)  = 0, 
\label{eq:2.2.20}
\end{equation}
with the generalized derivative $D_a \equiv \partial_a - \frac{6}{a}$. 
Directly from eq. (\ref{eq:2.2.5}), (\ref{eq:2.2.6}), (\ref{eq:2.2.7}), 
(\ref{eq:2.2.8})   we obtain 
\begin{equation}
B = a^3 h(\bar\phi) (1 + \bar\phi\phi)^{\frac{1}{2}} 
e^{3\sigma^2 a^2}~,D =   a^3 k(\phi) (1 + \bar\phi\phi)^{\frac{1}{2}} 
e^{- 3\sigma^2 a^2}~,
\label{eq:2.2.21b}
\end{equation}
where  
$h, k$ are anti-holomorphic and holomorphic 
functions of $\phi, \bar\phi$, respectively. 
Using again $r^2 = \bar\phi\phi$ and $\phi = r e^{i\theta}$, we obtain  
from eq. (\ref{eq:2.2.5})--(\ref{eq:2.2.8}) that 
\begin{eqnarray}
(1 + r^2) \frac{\partial \, B}{\partial \, r} 
- i\frac{1 + r^2}{r} \frac{\partial \, B}{\partial \, \theta} + rB 
& = & 0~, \label{eq:2.2.40} \\ 
(1 + r^2) \frac{\partial \, D}{\partial \, r} 
+ i\frac{1 + r^2}{r} \frac{\partial \, D}{\partial \, \theta} + rD
& = & 0~. \label{eq:2.2.41}
\end{eqnarray}
Multiply eq. (\ref{eq:2.2.40}) by $D$ and eq. (\ref{eq:2.2.41}) by 
$B$. Then divide by $1 + r^2$. Their subtraction 
eventually leads to  
\begin{equation}
 \frac{\partial (\, B \cdot D)}{\partial  \theta} 
- {i}{r} \left(\frac{\partial \, B}{\partial r} D 
- \frac{\partial \, D}{\partial \, r} B \right)
   =  0~. \label{eq:2.2.42}
\end{equation}

The  general  quantum state corresponding to a $k=1$ FRW supersymmetric 
model with a scalar supermultiplet is now given by 
\begin{eqnarray}
\Psi & =  & c_1 r^{\lambda_1} e^{-i\lambda_1 \theta} 
e^{- 3\sigma^2 a^2}  + 
c_3 a^3 r^{\lambda_3} e^{-i\lambda_3 \theta}(1 + r^2)^{\frac{1}{2}} 
e^{ 3\sigma^2 a^2}
  \psi^{C} \psi_{C} \nonumber \\ & + &  
c_4 a^3 r^{\lambda_4} e^{i\lambda_4 \theta}(1 + r^2)^{\frac{1}{2}} 
e^{- 3\sigma^2 a^2} 
\chi^{C} \chi_{C} + 
c_2 r^{\lambda_2} e^{i\lambda_2 \theta}
e^{3\sigma^2 a^2}
 \psi^{C} \psi_{C} \chi^{D} \chi_{D} ~,
\label{eq:2.New1}
\end{eqnarray}
where 
  $\lambda_1$...$\lambda_4$  
and $c_1$...$c_4$ are constants. 
Notice the explicit form of $A, B, D, E$ in (\ref{eq:2.New1}) 
in contrast with eq. (\ref{eq:2.2.20ba}), (\ref{eq:2.2.21b}) and ref. 
\cite{A9}-\cite{A10}, \cite{A16}-\cite{A18}, \cite{A19a,A23}. 
If we had use $\phi = \phi_1 + i\phi_2$ then the corresponding 
first-order differential equations  would lead to 
$A = d_1 e^{-3\sigma^2 a^2} e^{k_1(\phi_1 - i\phi_2)}$, 
$B = d_3 e^{3\sigma^2 a^2} (1 + \phi_1^2 + 
\phi^2_2) e^{k_3(\phi_1 - i\phi_2)}$,
$B = d_4 e^{-3\sigma^2 a^2} (1 + \phi_1^2 + 
\phi^2_2) e^{k_4(\phi_1 + i\phi_2)}$,
$E = d_2 e^{3\sigma^2 a^2} e^{k_2(\phi_1 + i\phi_2)}$.

As far as a  
generalization of relation 
(\ref{eq:2.2.20a}) is concerned, 
the  bosonic coefficients present in 
eq. (\ref{eq:2.New1}) satisfy  attractive relations
in a 3-dimensional minisuperspace:
\begin{eqnarray}
\frac{\partial (A\cdot E)}{\partial a} + 
\frac{\partial (A\cdot E)}{\partial \theta} - ir  \left(
\frac{\partial E}{\partial r} A - \frac{\partial A}{\partial r} E\right) & = &  0,
\label{eq:2.New22a} \\
D_a (B\cdot D)  + 
\frac{\partial (\, B \cdot D)}{\partial \, \theta } 
-ir  \left(\frac{\partial \, B}{\partial \, r} D 
- \frac{\partial \, D}{\partial \, r} B \right)
   & = &   0~.
\label{eq:2.New22b}
\end{eqnarray}
However, the presence of 
the terms 
$ ir  \left(
\frac{\partial E}{\partial r} A - \frac{\partial A}{\partial r} E\right)
$ and $
ir  \left(\frac{\partial \, B}{\partial \, r} D 
- \frac{\partial \, D}{\partial \, r} B \right)$ in
 eq. (\ref{eq:2.New22a}) and 
(\ref{eq:2.New22b}), respectively, 
clearly prevent us from obtaining  
 conservation  equations of the type  $\nabla \cdot J = 0$. 
The reason can be identified with  the 
variable $\theta$  no longer being a cyclical 
coordinate when supersymmetry is present\footnote{In fact, 
neither are 
$r$ or $\phi_1, \phi_2$ (defined from 
$\phi = \phi_1 + i\phi_2$) but in the 
non-supersymmetric description \cite{Khala} 
only $r$ is non-cyclical.} (see eq. 
(\ref{eq:2.2pitheta}) below).
To understand this argument, 
let us consider a FRW model with complex scalar fields in 
non-supersymmetric quantum cosmology 
(see e.g. ref. \cite{Khala}). 
The corresponding action  implies that the 
 conjugate momentum $\pi_\theta \sim   r^2 a^3 
\frac{\partial \, \theta}{\partial \, t}$ 
is a constant and $\theta$ constitutes a cyclical 
coordinate. However, in the corresponding 
supersymmetric scenario \cite{A22} there are  terms in the 
action that do {\it not} allow 
$\theta$ to be a cyclical coordinate. So, $\pi_\theta$ would not be a 
constant. And this will 
imply the absence of satisfactory
conserved currents.

In fact, the canonical  momenta 
conjugate to $r$ and $\theta$ 
take the following form:
\begin{eqnarray}
\pi_r &= & 2 \frac{\partial \, r}{\partial \, t} 
\frac{\sigma^2 a^3}{(1 + r^2)^2} - \frac{\sigma^2 a^3 e^{-i\theta}}
{\sqrt{2}(1 + r^2)^2}
3 n_{AA'} \chi^A \bar\psi^{A'}
+ \frac{\sigma^2 a^3 e^{i\theta}}
{\sqrt{2}(1 + r^2)^2}
3 n_{AA'} \chi^{A'} \psi^{A} \nonumber \\
 & - & \frac{\sigma^2 a^3 e^{-i\theta}}{\sqrt{2}(1 + r^2)^2} 
\chi^A\psi_{0A}  
 - 
\frac{\sigma^2 a^3 e^{i\theta}}{\sqrt{2}(1 + r^2)^2} 
 \bar\chi_{A'}\bar\psi_{0}^{A'} ~, 
\label{eq:2.2.pir}
\end{eqnarray}
\begin{eqnarray}
\pi_\theta & = &
\frac{2\sigma^2}{(1 + r^2)^2} r^2 a^3 
\frac{\partial \, \theta}{\partial \, t} 
+ 
\frac{5 \sigma^2 r^2 a^3}{\sqrt{2}(1 + r^2)^3}
n^{AA'} \bar\chi_{A'}\chi_A -   \frac{3 \sigma^2 r^2 a^3}{\sqrt{2}(1 + r^2)}
n^{AA'} \psi_A \bar\psi_{A'}\nonumber \\
& + & \frac{i r\sigma^2 a^3 e^{-i\theta}}
{\sqrt{2}(1 + r^2)^2}
3 n_{AA'} \chi^A \bar\psi^{A'}
+ \frac{i r \sigma^2 a^3 e^{i\theta}}
{\sqrt{2}(1 + r^2)^2}
3 n_{AA'} \chi^{A'} \psi^{A} \nonumber \\
 & +  & \frac{i r \sigma^2 a^3 e^{-i\theta}}{\sqrt{2}(1 + r^2)^2} 
\chi^A\psi_{0A}  
 - 
\frac{i r \sigma^2 a^3 e^{i\theta}}{\sqrt{2}(1 + r^2)^2} 
 \bar\chi_{A'}\bar\psi_{0}^{A'} ~.
\label{eq:2.2pitheta}
\end{eqnarray}
The relevant   point 
for our argument is that
equations  (\ref{eq:2.2.pir}) and  (\ref{eq:2.2pitheta}) 
directly prevent us to obtain relations like $\nabla J = 0$. To see this, 
notice the last four terms in equations  (\ref{eq:2.2.pir}) and  (\ref{eq:2.2pitheta}).
The origin of these specific terms can be traced back 
to the action of N=1 supergravity which implies that 
$\theta$ is no longer a cyclical variable. 

In addition, 
equations  (\ref{eq:2.2.pir}) and  (\ref{eq:2.2pitheta}) and the 
terms just mentioned  are basically translated into the last two terms 
present in
equations  (\ref{eq:2.New22a}) and 
(\ref{eq:2.New22b}), which are obtained 
from the supersymmetry constraints $S_A \Psi = 0$ and 
$\bar S_A \Psi = 0$. This follows  
directly  from the usual Hamiltonian 
 ${\cal H} \sim p\dot{q} - L$,
which involves 
a term $\sigma^2 a^3 \left[ \left(\frac{\partial \, r}{\partial \, t}\right)^2 + ir^2 
\left(\frac{\partial \, \theta}{\partial \, t}\right)^2 \right] $. 
But at this point we may emphasize as well the following. 

It is precisely the last two terms in each of the equations  (\ref{eq:2.2.pir}) and 
 (\ref{eq:2.2pitheta})  that will 
 allow 
us to obtain explicitely the contributions of the kinetic terms 
of $r$ and $\theta$ in 
  the supersymmetry constraints. These are then read 
from the coefficients of the Lagrange 
multipliers 
 $\psi^A_0,\bar\psi^{A'}_0$ in the 
Hamiltonian $\cal H$. 
But the last {\em four}  terms in both eq. 
(\ref{eq:2.2.pir}) and (\ref{eq:2.2pitheta}) 
(which also include the ones with 
$\psi^A_0, \bar \psi^{A'}_0$)  are also a 
direct 
consequence of  the  terms in the 
  action \cite{8} that imply $\theta$ not being a cyclical coordinate. 
Thus, the fact that the coordinate 
$\theta$ is no longer being a cyclical coordinate
can be interpreted as  inherited from local supersymmetry, which is 
 now a feature of the reduced model.
This can be summarized  as follows: a relation as $\nabla \cdot J=0$
{\em cannot} be sensibly defined
 due to 
 the   absence  of cyclical coordinates,  which are ultimately due to the presence of 
supersymmetry. A similar situation\footnote{The author is 
grateful to S. Kamenshchik for having pointed out this to him.} 
 would occur in usual quantum cosmology 
with a matter Lagrangian taken from the Wess-Zumino model, 
due to the non-trivial interaction with fermion fields.

\section{Bianchi class-A models}

\indent

  Bianchi class-A models obtained within  
pure N=1 supergravity are analysed in this section, closely 
following ref. 
 \cite{A12}   (see also ref. \cite{A21,A20,rev,FPDD}). 
A left-invariant basis \cite{Ryan} is employed, 
where the spatial metric $h_{ij}$  
can be 
expanded as  
$h_{ij} = h_{pq}(t) E^p_{~i} E^q_{~j}.
$ 
The left-invariant invariant basis $E^p_{~i}$ further satisfy
$\partial_{[i}E^p_{~j]} = C^p_{~qr} E^q_{~i} E^r_{~j}$.  
$C^p_{~qr} = m^{pt} 
\varepsilon_{tqr}
$ are the structure constants of the Bianchi group
which allow to identify each Bianchi model. 
In this  basis   the tetrad 
$e^{AA'}_{~~p}(t)$
is 
time-dependent only. Notice that 
$h_{pq} = - e_{AA'p} e^{AA'}_q$. In addition, we have 
$\psi^A_{~0} = \psi^A_{~0}(t), \psi^A_{~i} = \psi^A_{~p}(t) E^p_{~i}  
$ 
with similar restrictions imposed on $\bar\psi^{A'}_{~\mu}$. 
An important feature of Bianchi class-A models is that 
 anisotropic gravitational degrees of freedom are now present. Hence, more 
gravitino  modes are allowed to be included. 
In such a way a more realistic insight on 
 the full theory of N=1 supergravity can be obtained.

Let us  focus 
 on the question if 
conserved currents can be constructed 
in these supersymmetric 
models. 
As we mentioned before, our approach 
 bears  significant differences as 
far as ref. \cite{OOcc}-\cite{A4} 
are concerned. The Lorentz constraint seems  either absent in ref. 
\cite{A1}-\cite{A4} or 
truncated to some extent ref. \cite{OOcc}. Such absence  may  enforce
additional restrictions, either in the spectrum of solutions or 
even in 
the 
general validity of the results. 
Our approach involves instead a fermionic differential operator 
representation and the {\em 
complete} Lorentz constaints. This is of some 
relevance since    the 
wave functional ought  to be a Grassman-algebra-valued expression and not 
just a wave function arranged as a vector \cite{OOcc}-\cite{A4}.

The  Lorentz invariance implies that  
  $\Psi$ can only contain fermionic terms with an even number 
of $\psi^A_{~p}$. Thus, we can decompose (and not restrict!) 
$\Psi$ into fermionic parts of 
zeroth (bosonic), quadratic, quartic up to sixth (fermionic filled) 
order, formally  denoted by $\psi^0, \psi^2, \psi^4, \psi^6$. 
Moreover, we  employ the   general decomposition of $\psi^A_{~i}$ as   
$\psi^A_{~~B B'} = e_{B B'}^{~~~~i} \psi^A_{~~i} $ and 
$
 \psi_{A B B'} = - 2 n^C_{~~B'} \gamma_{A B  C} + {2 \over 3} \left( \beta_A
n_{B B'} + \beta_B n_{A B'} \right) - 
2 \varepsilon_{A B} n^C_{~~B'} \beta_C$, 
where the $ \gamma_{A B C} = \gamma_{(A B C)} $ and $\beta^A$ 
fields denote the spin-$\frac{3}{2}$ and $\frac{1}{2}$ modes of 
the gravitino, respectively. The wave function of the universe 
can then be symbolically written  as 
\begin{equation}
\Psi = A (h_{pq}) + \psi^2 + \psi^4 + F(h_{pq}) 
\beta^A\beta_A\left(\gamma^{BCD}\gamma_{BCD}\right)^2~,
\label{eq:3.14}
\end{equation}
where $A$ and $F$ are the coefficients of the bosonic ($\psi^0$) and 
fermionic filled ($\psi^6$) sectors; the middle sectors $\psi^2$ and 
$\psi^4$ require  further elements and will be discussed  later in this 
section.

The supersymmetry constraints for a generic Bianchi 
class-A minisuperspace have  the form \cite{A12}
\begin{equation}
 S_{A} =  \sigma m^{pq} e_{AA'p} \bar\psi^{A'}_q 
- \frac{1}{2}i\kappa^2 \left[
(1-s) p_{AA'}^p \bar\psi^{A'}_p + s \bar\psi^{A'}_p p_{AA'}^p~\right],
\label{eq:3.2}
\end{equation}
together with its Hermitian conjugate, 
where we take $\kappa^2 = 8\pi$.
The parameter $s$ represents the ambiguity 
of operator ordering, which comes from noncommutativity of 
$\psi^A_p, \bar\psi^{A'}_q, p_{AA'}^r$.
$p_{AA'}^q$ constitutes the conjugate momenta to $e^{AA'}_q$. 
Quantum mechanically, the $\bar S_{A'}$ 
supersymmetry constraint takes the form 
\begin{equation}
\zeta m^{pq} e_{AA'p} \psi^A_{~q}  +   \psi^A_{~p}
\frac{\partial}{\partial 
e^{AA'}_p} + s \psi^A_{~p}e_{AA'}^p = 0, \label{eq:3.11} \end{equation}
where $\zeta = \frac{2\sigma}{\hbar\kappa^2}$ and $S_A$ is just the 
Hermitian conjugate. 
We have chosen the  following representation \cite{4,A12}
\begin{eqnarray}
\bar\psi^{A'}_{~p} & = &   -i\hbar 
D^{AA'}_{~~qp} \frac{\partial}{\partial \psi^A_{~q}}~, 
\\ 
 p_{AA'}^p  & = & -i\hbar \frac{\partial}{\partial e^{AA'}_{~~p}}
+ \frac{1}{2} i\hbar \sigma \varepsilon \psi^A_{~q} 
D^{BA'}_{tr}\frac{\partial}{\partial \psi^B_{~t}},
\end{eqnarray} 
 where 
\begin{equation}
D^{AA'}_{~~pq} = 
\frac{i}{\sigma \sqrt{\det h}}  h_{pq}n^{AA'} + 
\frac{\varepsilon_{pqr}}{\sigma} e^{AA'r}
\end{equation}
and $\sigma$ is the volume of the hypersurface of 
quantization.

It can be checked from equations (\ref{eq:3.11}), its  Hermitian 
conjugate and  the wave function 
(\ref{eq:3.14}) 
that the bosonic 
coefficients $A$ and $F$ in (\ref{eq:3.14}) satisfy, 
 respectively, the equations 
\begin{eqnarray}
\left[2\frac{\partial}{\partial \,h_{pq}} - \zeta m^{pq}   -s h^{pq}
\right] A
 = 0~,\label{eq:3.15}\\
\left[2\frac{\partial}{\partial \,h_{pq}}+ \zeta m^{pq}    +s  h^{pq}
\right] F
 = 0~.\label{eq:3.16}
\end{eqnarray}
The corresponding solutions\footnote{The solution  (\ref{eq:3.16a}) is 
{\em different}  from the corresponding expression
in ref. \cite{A12}. The extra $h$ factor present in eq. (35) of ref. 
\cite{A12} {\it cannot} 
 be there though. 
The explicit form of (\ref{eq:3.16a}) can be 
obtained through a 
fermionic Fourier transformation \cite{4,A12}
\begin{equation}
\bar\Psi (e^{AA'}_{~~p}, \bar\psi^{A'}_{~q}) 
= D^{-1} (e^{AA'}_{~~p}) \int \Psi( e^{AA'}_{~~p}, \psi^A_{~q})
e^{-\frac{i}{\hbar}C^{~~pq}_{AA'}\psi^A_{~p}\bar\psi^{A'}_{~q}}
\Pi_{E,r}d\psi^E_{~r}~,
\label{eq:3.jap1}
\end{equation}
where 
$
C_{AA'}^{~~pq} = - \sigma \varepsilon^{pqr}e_{AA'r}~, ~
D (e_{AA'p}) = \det \left(-\frac{i}{\hbar} C_{AA'}^{~~pq}\right)$. 
This gives us the wave function in the representation 
$\bar\Psi (e^{AA'}_{~~p}, \bar\psi^{A'}_{~q})$ and 
leads to a factor of $h^{-1}$ (via $
D^{-1} (e^{AA'}_{~~p})$). 
In particular, it   intertwines $F$ in (\ref{eq:3.14}) 
with $\bar A$,  whose equation is substantially easier to derive. 
But the inverse transformation does {\it not} involve a factor of $h$; 
see ref. \cite{4} for the reasons of this asymmetry.} are 
\begin{eqnarray}
A & = & c_1 (\det h_{pq})^{\frac{s}{2}} e^{\frac{1}{2}\zeta 
m^{pq}h_{pq}}~, \label{eq:3.15a} \\
F & = & c_2   (\det h_{pq})^{-\frac{s}{2}} e^{-\frac{1}{2}\zeta 
m^{pq}h_{pq}}~, \label{eq:3.16a}
\end{eqnarray}
where $c_1, c_2$ are constants. 
Let us now multiply eq. (\ref{eq:3.15}) by F and 
eq. (\ref{eq:3.16}) by A. We find after  adding them that  
\begin{equation}
\frac{\partial ( A\cdot F)}{\partial \, h_{pq}}   = 0~.
\label{eq:3.17}
\end{equation}
As expected, eq. (\ref{eq:3.17}) 
constitutes the generalization of eq.  (\ref{eq:2.2.20a}). 
Moreover, eq. (\ref{eq:3.17}) also represents  
  the decomposition of
the result 
present in ref. \cite{A4} concerning a (positive-definite) 
probability 
amplitude conservation. However,  it is viewed in this section  
 within a fermionic differential operator representation 
\cite{4,rev}.

Let us now consider the 
middle fermionic sectors of $\Psi$. 
As far as these   sectors are concerned, consistency 
can only be achieved  (see ref. \cite{A21,A20})  
from   the use of the Hamiltonian 
constraint  derived from the Dirac bracket $\left[ S_A, \bar S_{A'}\right]_D$.
However, this {\em back to basics}  \cite{A20a} 
procedure clearly move   us  
  from the purpose of using solely  
the first-order differential equations generated by  
 the supersymmetry constraints. 
Any  conserved currents (and positive-definite 
probability densities)  must then be addressed within 
the  Wheeler-DeWitt-type equations 
obtained in 
\cite{A21,A20}. 

The situation concerning the full theory  of N=1 supergravity 
is even more helpless. As we mention in the introduction, 
a conservation equation of the form $\nabla_a  J^a = 0$ 
can be derived   in general relativity with the assistance 
of the Wheeler-DeWitt equation \cite{G26a}. 
But this is achieved {\em without} 
making any assumptions about  the space-time geometry. 
 With respect to N=1 supergravity
the $\bar S_{A'} \Psi = 0 $ constraint reads \cite{4}
\begin{equation}
\left(\varepsilon^{ijk} e_{AA'i} {}^{3s}D_j\psi^A_{~k}
\right) \Psi - \frac{1}{2}
\kappa^2 \psi^A_{~i} \frac{\delta \Psi}{\delta e^{AA'}_{~~i}} 
= 0 ,
\label{eq:full1}
\end{equation}
while  the 
$S_A \Psi = 0$ constraint in the $\bar\Psi (e_{AA'i}, \bar\psi^{A'}_{~i})$ 
representation  
is given by 
\begin{equation}
\left(\varepsilon^{ijk} e_{AA'i} {}^{3s}D_j\psi^{A'}_{~k}
\right) \bar \Psi + \frac{1}{2}
\kappa^2 \bar\psi^{A'}_{~i} \frac{\delta \bar\Psi}{\delta e^{AA'}_{~~i}} 
= 0 ~.
\label{eq:full2}
\end{equation}
In the homogenous case, the $\psi^A_{~i}$ and $\bar\psi^{A'}_{~i}$ 
can be arbitrarily  chosen and hence cancelled out 
in eq. (\ref{eq:full1}) and (\ref{eq:full2})
\cite{A6,A12,A14}. 
Furthermore,  we  can   use either of 
the  representations $\Psi$ or $\bar\Psi$:  
  their 
bosonic coefficients are related by a Fourier transformation 
generalizing (\ref{eq:3.jap1}). But the 
inhomogenous case is  clearly different. The gravitinos {\it cannot} 
be cancelled out. Moreover,  $\Psi$ can only have 
states with {\it infinite} fermion number 
 in full N=1 supergravity \cite{infi}.

\section{Our message}

\indent 

The purpose of this paper was to 
investigate if   conservation equations  
  \cite{Kaku} of the type  $\nabla \cdot J = 0$ 
could  be sensibly defined in supersymmetric minisuperspaces. 
In section 2 we considered $k=+1$ FRW models  
with   and without supermatter 
  in the form of scalar supermultiplets 
$\left(\phi, \bar\phi; \chi_A, \bar\chi_{A'}\right)$. 
Bianchi class-A models in  pure N=1 supergravity 
as well as the full theory 
were  discussed in section 3.
We  restricted ourselves to the  
first-order 
differential equations derived from  
 the Lorentz and supersymmetry constraints.  
Moreover, we
employed here   a  
differential operator representation
 for the  fermionic variables.
This constitutes the correct  approach 
due to 
 the existence of second--class constraints 
and subsequent  Dirac brackets. 
These  then imply that  the  fermionic variables and their Hermitian conjugates are
 intertwined within  a canonical coordinate-momentum relation. 
The ``square-root'' structure present in the 
algebra of constraints of 
N=1 supergravity was the main motivation for 
our study. Namely, the fact that 
the above mentioned  
first-order differential equations act relatively 
to the Wheeler-DeWitt equation in a way similar to the 
standard procedure in quantum field theory 
relating  the Klein-Gordon and Dirac 
equations \cite{Kaku}.

In the supermatter case our results were twofold.
First, we showed how the explicit dependence of the 
wave functional $\Psi$ on $\phi, \bar\phi$ could be 
brought about. This was done 
by introducing the 
the transformation 
$\phi = r e^{i\theta} = \phi_1 + i\phi_2$ 
  {\it directly} in the supersymmetry constraints.
In particular, the no-boundary wave function has now been 
adequately identified, in contrast with previous comments 
in ref. 
\cite{A16}-\cite{A18}, \cite{A19a,A23}.

Our second result also followed from the use of this transformation.
In fact,  the variable $\theta$ is no longer a cyclical coordinate 
if supersymmetry is present. This should be 
 contrasted  with 
the situation    in plain quantum cosmology. 
Basically, $\theta$ being a non-cyclical coordinate 
is caused by the presence of specific   terms in the 
 momenta  $\pi_r$ and 
$ \pi_\theta$,   which
lead to the expressions  
$ \frac{\partial (A\cdot E)}{\partial a} + 
\frac{\partial (A\cdot E)}{\partial \theta} - ir  \left(
\frac{\partial E}{\partial r} A - \frac{\partial A}{\partial r} E\right) = 0$ and 
$ D_a (B \cdot  D) + 
\frac{\partial (\, B \cdot D)}{\partial \, \theta } 
-ir  \left(\frac{\partial \, B}{\partial \, r} D 
- \frac{\partial \, D}{\partial \, r} B \right) = 0 $ 
( see eq. (\ref{eq:2.New22a}) and  (\ref{eq:2.New22b})). 
And these expressions clearly prevent us 
from obtaining a relation 
like  $\nabla \cdot J = 0$ {\em directly} from the supersymmetry 
constraints.
But 
among those specific 
 terms in $\pi_r$ and $\pi_\theta$ (that prevent 
$\nabla J = 0$ to be obtained), we can also identify the ones  
 which are necessary to 
retrieve the supersymmetry constraints in a usual 
canonical formalism. Hence, 
the absence of satisfactory conserved currents may be ultimately 
related to the presence of 
supersymmetry in our supermatter model. Furthermore, 
our   results    provide a dissimilar   perspective 
with regard to 
 ref. \cite{A17a}. There,  a particular square-root of a  
{\em non}-supersymmetric 
FRW model with complex scalar fields was used. 
Our model is sustained instead by a ``square-root'' 
quantization inherited from N=1 supergravity.

Concerning homogeneous models in pure N=1 
supergravity, we obtain 
a rather simple conservation equation  
for the FRW case. With respect to Bianchi models, 
our results are 
summarized in eq. (\ref{eq:3.17}). This expression 
 represents essentially 
the conservation of the 
probability amplitude mentioned in ref. \cite{A4} but 
viewed here within our canonical approach. Namely, 
by employing a differential operator representation 
for the fermionic variables. This should be 
compared   with ref. \cite{OOcc}-\cite{A4}, where 
the Lorentz constraint is   either absent or   
truncated. Moreover, the supersymmetric wave function
in  \cite{OOcc}-\cite{A4} 
is 
arranged in a vector, leading to universe---anti-universe states.

\vspace{1cm}

Overall, our message in this paper is
that {\em generic} conserved currents do not 
seem feasible  to obtain {\em directly} from the supersymmetry 
constraints equations. 
Only for very simple scenarios does this becomes 
possible. 
Otherwise, conserved currents (and consistent probability 
densities) may  only be    obtained  upon the use 
of 
subsequent Wheeler-DeWitt--like equations. These are 
derived through  the associated 
 supersymmetric algebra of constraints.

In our view, the fundamental reason for our conclusions 
 is related with 
the following.  A physical supersymmetric 
wave functional $\Psi$ 
takes values in a Grassman algebra. Such algebra  
is formed by complex linear combinations of 
products of anti-commuting elements such as 
the gravitino $\psi^A_i$. 
Hence, 
$\Psi \left[e^{AA'}_\mu, \psi^A_\mu, \bar\psi^{A'}_\mu; 
\phi, \bar\phi, \chi^A, \bar\chi^{A'}\right]$ embodies more than a  
  wave function arranged as a vector and satisfying a Dirac-like equation 
(see ref. \cite{OOcc}-\cite{A4}).
Furthermore,  the  first-order differential equations  
derived from the supersymmetry, Lorentzian and Grassmanian-valued 
$\Psi$   constitute  more than a simple 
 set of conditions. They rather 
represent the action of the supersymmetry constraints 
on {\em different}  fermionic representations of $\Psi$, related by a 
(coordinate--momentum) fermionic Fourier transformation 
\cite{4,A8}--\cite{A11}.

Finally, we included two appendices. One contains 
additional  comments on quantum FRW closed cosmologies with 
complex scalar fields in the 
presence of supersymmetry. 
The other appendix initiates a discussion on  the retrieval of classical properties 
from a supersymmetric quantum cosmological 
scenario.

\appendix

\section{Supersymmetric quantum FRW models with complex 
scalar fields}

\indent

In this appendix we will compare some of the features of 
supersymmetric quantum cosmological FRW models in the 
presence of complex scalar fields with the corresponding situation 
in non-supersymmetric quantum cosmology. To begin with, let us mention that 
FRW quantum cosmological models with 
complex scalar fields can be found  in, e.g.,  
ref. \cite{Khala}. A characteristic  method employed  in 
\cite{Khala} was the use of the following transformation for the 
complex scalar field:  
$\phi \rightarrow r e^{i\theta}$. This allowed 
to identify $\theta$ as cyclical coordinate and hence a 
constant conjugate momentum: $\pi_\theta \sim   r^2 a^3 
\frac{\partial \, \theta}{\partial \, t}$. 

In this paper we also investigated  
closed FRW models with 
complex scalar fields but in the context of supersymmetric 
quantum cosmology. The definition  $\phi = r e^{i\theta}$ was 
also employed but {\em directly}
in the supersymmetry constraints. 
Notice that in ref. \cite{A10a,A10} this  was only employed 
in subsequent Wheeler-DeWitt--type equations. 
One of the advantages of 
in substituting $\phi = r e^{i\theta}$ in the supersymmetry 
constraints is to obtain the 
explicit dependence of $\Psi$ on $\phi, \bar \phi$. 
This was missing in ref. \cite{A9}-\cite{A10}, 
\cite{A16}-\cite{A18}, \cite{A19a,A23}. 

In addition, this appraoch also allowed our results to be compared with 
the ones present in \cite{Khala}. 
In fact, the bosonic coefficents in expression  (\ref{eq:2.New1})
correspond 
to particular solutions that can be obtained in   the framework of 
ref. \cite{Khala}. Namely,  {\it if} a specific factor ordering for 
$\pi_a, \pi_r, \pi_\theta$ is used in the Wheeler-DeWitt equation 
together with  a judicious choice of integration constants.  The point is that 
the  constraint (\ref{eq:2.13}), its Hermitian 
conjugate, the 
Lorentz constraint  and expression 
(\ref{eq:2.15}) imply $\frac{\partial A}{\partial \phi} =0$ 
and $\frac{\partial E}{\partial \bar\phi} =0$ 
(i.e., eqs. (\ref{eq:2.2.20d}), (\ref{eq:2.2.20e})) and hence the solutions 
you found here for $A$ and $E$. Let us now see how such expressions 
can be obtained from the Hamiltonian cosntraint It is then important to notice that the Hamiltonian 
constraint includes  a
bosonic term 
$\pi_\phi \pi_{\bar\phi} \sim 
(\pi_r -i\pi_\theta)(\pi_r + i\pi_\theta)$. 
Quantum mechanically, we can write it  as 
$\frac{\partial^2 }{\partial r^2}   +  
 \frac{1}{r^2}\frac{\partial^2 }{\partial \theta^2}.$
This was in fact the choice made in ref. \cite{Khala}.
But notice that we could also have 
$ (\pi_r -i\pi_\theta)(\pi_r + i\pi_\theta)$ as 
$\left(\frac{\partial }{\partial r}   -  
 i\frac{1}{r}\frac{\partial }{\partial \theta}\right)
 \left(\frac{\partial }{\partial r}   +   
 i\frac{1}{r}\frac{\partial }{\partial \theta}\right) 
$ which is different from $  
 \frac{\partial^2 }{\partial r^2}   +  
 \frac{1}{r^2}\frac{\partial^2 }{\partial \theta^2}$. 
Hence, the presence of supersymmetry 
induces a particular set of solutions for $\Psi$, which 
can only be obtained from the Wheeler-DeWitt equation if 
 a particular factor ordering for 
the canonical momenta 
is selected in the Hamiltonian constraint.
As a consequence, 
specific and explicit 
exact solutions (say, $e^{-3\sigma^2 a^2} r^\lambda e^{i\lambda\theta}$) 
can be found in the 
gravitational and matter sectors.

A comparison between quantum FRW models with and without supersymmetry 
can be further enhanced by mentioning the following. Firstly, it is 
interesting to remark that when 
supersymmetry 
is absent, the FRW minisuperspace is formed 
by two ``independent'' sectors represented by 
the variables $\{a,r\}$ and $\{\theta\}$, respectively (see 
ref. \cite{Khala}). The 
only apparent influence of $\theta$ is the presence of the constant 
values of $\pi_\theta$ in the  the $\{a,r\}$ sector
of the 
Wheeler-DeWitt equation. But if 
supersymmetry is present the situation changes considerably. 
The variables $\{a,r,\theta\}$ become all intertwined 
and the analysis is less simple. 
Furthermore, 
the constants of separation $\lambda_1, ..., \lambda_4$ seem
to correspond to $i\equiv\sqrt{-1}$ times 
the eigenvalues of 
$\pi_\theta$ permited 
 in \cite{Khala}.

Finally, let us mention that 
the no-boundary wave function  
corresponds to the bosonic coefficient $A$. We stress that only  
by employing $\phi = r e^{i\theta} = \phi_1 + 
i\phi_2$ can such claim    become fully justified. 
The exponential factors  $e^{\pm 3\sigma^2 a^2}$  are 
to be viewed  as $e^{\pm I}$, where $I$ is {\it Euclidean} action 
for a classical solution {\em without} matter outside or inside a 
three-sphere with radius $a$ (see ref. \cite{A8,A9}). 
In the absence of matter 
the Hartle-Hawking state \cite{12}   is therefore 
given by $\Psi_{HH} = \psi_A\psi^A e^{3\sigma^2 a^2}$. 
The solution $\Psi = e^{-3 \sigma^2 a^2}$ bears quantum 
wormhole properties \cite{A10a,A10,A19a,A23,13}.
Furthermore, we get the expressions 
$A = d_1 e^{-3\sigma^2 a^2} e^{c_1(\phi_1 - i\phi_2)}$ and   
$E = d_2 e^{3\sigma^2 a^2} e^{c_2(\phi_1 + i\phi_2)}$
from the decomposition 
$\phi= \phi_1 + i\phi_2$.  
For each of them the  constant $k$
(as defined from 
the equations (9)-(23) of ref. \cite{13})
has to be $k = d_{1(2)}^2 - d_{1(2)}^2 = 0$. 
This means that the 
scalar flux associated with   
{\em two}  independent massless 
scalar fields $\phi_1$ and $\phi_2$  is now absent.
Consequently, there is also no lower 
bound for $a$. 
Such result is quite curious. In fact, it is not apparent  how the 
solutions  for $A$ and $E$ could 
 represent a wormhole connecting 
two asymptotic regions. For a related discussion 
concerning  the  retrieval of wormhole states  see ref. \cite{A23}.

\vspace{1cm}


\section{Can classical properties  be retrieved from 
supersymmetric quantum cosmologies? }

\indent

In this appendix we present some comments concerning the 
retrieval of classical features in 
supersymmetric quantum cosmology. The main reason is that 
the topic  of conserved 
currents (and positive-definite probability amplitudes) 
is particularly relevant  in quantum cosmology and  closely related with  
  the emergence  of classical properties from 
quantum   models \cite{5}. Hence the question we raise here: 
would the  presence of supersymmetry introduce  any significative changes 
concerning the retrieval of classical properties in the framework of 
quantum cosmology?

An answer to this question is currently being sought after 
\cite{pmnew}. In order to understand the relevance of 
retrieving classical properties from supersymmetric quantum cosmologies, 
let us mention the following. 

 In plain   quantum cosmology,   conserved probability 
currents can be obtained by requiring 
the wave function of the 
universe to be  of the form $\Psi_{WKB}  \sim e^{iS}$.
As consequence, classical properties can emerge from $\Psi_{WKB}$. 
But what can be obtained from $e^I$?
When $\Psi$ is an exponential $e^I$
rather than an oscillatory  function, $I$   corresponds to the action of  an 
Euclidean rather than a Lorentz  geometry. This situation occurs when no
 matter is present, and 
the dominant saddle-point contribution to the 
path--integral    solution is a real Euclidean 
solution of the field equations \cite{swh31}. 
This conclusion holds for a variety of homogeneous models \cite{swh31}.
But the important fact to notice is that 
a  wave functional $e^{I}$ is {\em not} peaked around a set of Euclidean 
solutions. It fails to predict classical correlations  between 
bosonic coordinate and momenta. In contrast, a wave function 
$e^{iS}$  is peaked around a set of classical Lorentzian trajectories
\cite{5,kiefer}.

In supersymmetric 
quantum cosmology 
most of the solutions that have been found  include {\it 
only }the exponential of the Euclidean action
$e^{\pm I}$   \cite{rev,A5}-\cite{A18a}, \cite{A21,A20}.
This means that these solutions 
present in \cite{A21,A20} do {\em not} 
induce any classical Lorentzian geometry. 
Thus the current framework for supersymmetric Bianchi  models 
may require additional elements 
 in order to get 
oscillating  $e^{iS}$ solutions. Maybe that would 
provide the means to establish 
a 
relation  similar to  
$\nabla \cdot J =0, ~ 
J \sim e^{-I_{R}} |C^2| \nabla S$, 
which is
only valid in a minisuperspace approximation. 
However, solutions with oscilating properties 
were  
obtained in ref. 
\cite{A14,A15}
for a very simple FRW model with $\Lambda \neq 0$ and 
within pure N=1 supergravity.

Finally, the relation between supersymmetry and 
first-order differential equations may also provide another interesting 
perspective\footnote{The author is grateful to C. Kiefer 
for having pointed out this to him.}. Previous analysis 
based on the second-order Wheeler-DeWitt equation has shown a 
peculiar behaviour 
when constructing 
wave packets \cite{54} in quantum cosmology. Perhaps the general superposition 
present in the wave functional 
(\ref{eq:2.15}) may point out to alternative insights.

\vspace{1cm}

{\large\bf Acknowledgements}

\vspace{0.3cm}

This work was supported by   the 
 JNICT/PRAXIS--XXI Fellowship BPD/6095/95. 
The author 
is grateful to 
A. Yu. Kamenshchik and C. Kiefer 
 for useful  comments and discusions.
Correspondence with M. Asano is  acknowledged
as well as conversations with 
  R. Graham and  O. Obr\'egon which 
motivated some of the analysis discussed in this paper.
Finally, the author is grateful 
to an anonymous  for a critical reading of the manuscript and for pointing to 
improvements.

\end{document}